\documentclass[conference]{IEEEtran}

\usepackage{mdframed}
\usepackage{balance}
\usepackage{amssymb,bm,mathtools}
\usepackage{amsmath}
\allowdisplaybreaks
\usepackage{epsfig,subfigure,colortbl,psfrag,subfloat}
\usepackage{enumerate,array,multirow}
\usepackage{cite,url}
\usepackage{algorithmic}
\usepackage{algorithm,setspace}
\usepackage{framed}
\usepackage{threeparttable,supertabular}

\newcommand{\vect}[1]{{\mathbf{#1}}}
\newcommand{\mat}[1]{{\mathbf{#1}}}
\newcommand{\beqna}{\begin{IEEEeqnarray}{rCl}}
\newcommand{\eeqna}{\end{IEEEeqnarray}}

\renewcommand{\r}{\vect{r}}

\newcommand{\x}{\vect{x}}

\newcommand{\z}{\vect{z}}

\renewcommand{\H}{\mat{H}}
\newcommand{\I}{\mat{I}}

\newcommand{\U}{\mat{U}}

\newcommand{\W}{\mat{W}}

\newcommand{\bmat}{\begin{bmatrix}}
\newcommand{\bemat}{\end{bmatrix}}
\newcommand{\nn}{\nonumber}
\newcommand{\eref}[1]{(\ref{#1})}

\newcommand{\bmu}{\mat{\boldsymbol{\mu}}}

\newcommand{\bomega}{\mat{\boldsymbol{\omega}}}

\newtheorem{lem}{Lemma}

\newtheorem{rem}{Remark}

\def \beqi{\begin{IEEEeqnarray}{rcl}\IEEEyesnumber}
\def \eeqi{\end{IEEEeqnarray}}
\def \inum{\IEEEyessubnumber}

\def \beq {\begin{equation} }
\def \eeq {\end{equation} }
\def \beqn{\begin{eqnarray} }
\def \eeqn{\end{eqnarray} }

\begin{document}

\title{Opportunistic Multicast Scheduling for Unicast Transmission in MIMO-OFDM System}

\author{\IEEEauthorblockN{Peng Hui Tan, Jingon Joung, Sumei Sun}
\IEEEauthorblockA{Institute for Infocomm Research (I$^2$R), A$^\star$STAR,
Singapore\\
Email:\{phtan, jgjoung, sunsm\}@i2r.a-star.edu.sg}}

\maketitle

\begin{abstract}
We propose a multicast scheduling scheme to exploit content reuse when there is asynchronicity in user requests. A unicast transmission setup is used for content delivery, while multicast transmission is employed opportunistically to reduce wireless resource usage. We then develop a multicast scheduling scheme for the downlink multiple-input multiple-output orthogonal-frequency division multiplexing system in IEEE 802.11 wireless local area network (WLAN). At each time slot, the scheduler serves the users by either unicast or multicast transmission. Out-sequence data received by a user is stored in user's cache for future use. Multicast precoding and user selection for multicast grouping are also considered and compliance with the IEEE 802.11 WLAN transmission protocol. The scheduling scheme is based on the Lyapunov optimization technique, which aims to maximize system rate. The resulting scheme has low complexity and requires no prior statistical information on the channels and queues. Furthermore, in the absence of channel error, the proposed scheme restricts the worst case of frame dropping deadline, which is useful for delivering real-time traffic. Simulation results show that our proposed algorithm outperforms existing techniques by $17~\%$ to $35~\%$ in term of user capacity.

\end{abstract}
\begin{keywords}
Multicast scheduling, Lyapunov optimization, multicast precoding, WLAN network
\end{keywords}

\section{Introduction}\label{sec:intro}

\begin{figure}[!t]
\begin{center}
\psfrag{q}[rc][cc][.8][0]{\sf Queue 1}
\psfrag{p}[rc][cc][.8][0]{\sf Queue 2}
\psfrag{u}[lc][cc][.7][0]{\sf user (STA) 1}
\psfrag{v}[lc][cc][.7][0]{\sf user (STA) 2}
\psfrag{t}[cc][cc][.8][0]{\sf $t_1$}
\psfrag{r}[cc][cc][.8][0]{\sf $t_2$}
\psfrag{d}[cc][cc][1][0]{\sf $\cdots$}
\psfrag{e}[cc][cc][.8][0]{\sf D1}
\psfrag{c}[cc][cc][.8][0]{\sf D2}
\psfrag{a}[cc][cc][.8][0]{\sf D3}
\psfrag{f}[cc][cc][.8][0]{\sf D3}
\psfrag{g}[cc][cc][.8][0]{\sf D4}
\psfrag{h}[cc][cc][.8][0]{\sf D5}
\subfigure[\!\!\!\!\!]{%
\epsfxsize=0.28\textwidth \leavevmode
\epsffile{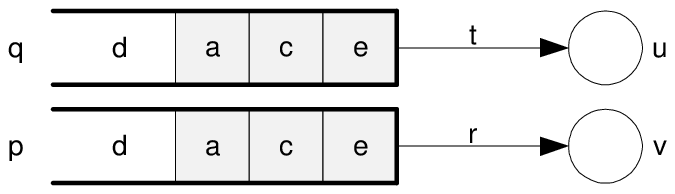}}\\
\subfigure[\!\!\!\!\!]{%
\epsfxsize=0.28\textwidth \leavevmode
\epsffile{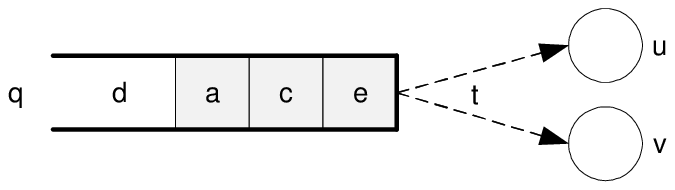}}\\
\subfigure[\!\!\!\!\!]{%
\epsfxsize=0.28\textwidth \leavevmode
\epsffile{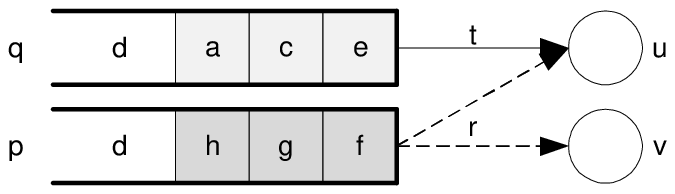}}
\caption{Illustration of (a) unicast. (b) multicast. (c) opportunistic multicast.}
\label{fig:f0}
\end{center}
\end{figure}

To provide satisfactory quality of service (QoS) for multimedia contents, efficient allocation of wireless resource is a necessity. Opportunistic scheduling is one of the most promising techniques. It has been observed that most of user requests are restricted to only a few very popular contents. For such scenario, multicast is an efficient mechanism for one-to-many transmissions over wireless channels \cite{Won09WC,Low10WC,Tsa11WC}. Contrary to a unicast, in which each user (or STA: station) is supported by an access point (AP) separately at each time slot $t_1$ or $t_2$ as illustrated in Fig. \ref{fig:f0}(a), multicast can support multiple users who request identical content simultaneously as illustrated in Fig. \ref{fig:f0}(b). Herein, users 1 and 2 belong to a multicast group, which requires message (data chunk, internet protocol packet, or data frame) D1, D2 and D3 from the AP. 
On the other hand, the user requests usually occur at different times, i.e., asynchronous  request. Hence, the AP has to fall back to unicast transmission and loses the exploitation of this content reuse feature. Another approach to deal with the opportunistic demand is harmonic broadcasting and its variants introduced in \cite{Juh97B,Cha07CSVT}. These schemes enable each user to start playback within a small delay from its request time. However, the allocation of wireless resource, i.e., scheduling in time slot, was not considered.





In this work, our goal is to develop efficient transmission and scheduling (i.e., time resource allocation) scheme to exploit the content reuse feature under the opportunistic requests. We refer to this transmission scheme as the opportunistic multicast, and illustrate it in Fig. \ref{fig:f0}(c). Herein, users 1 and 2 demand for identical content. Since unicast transmission is used, two queues are required at the AP. However, either unicast or opportunistic  multicast transmission is performed dynamically at each time slot. For example, if queue 1 is scheduled for transmission, AP sends message D1 to user 1 only, which is the same as the unicast transmission, as user 2 has already received D1. On the other hand, if AP selects queue 2 for transmitting message D3 to (intended) user 2 and if it also knows that user 1 requires D3 in the future, the AP switches from unicast to multicast transmission and sends D3 to both users 1 and 2. Once user 1 receive D3, it stores D3 in its own cache for future use\footnote{Our work is mainly inspired by the caching approach to deliver contents. In \cite{Gol13Arx,Ji13ISIT,Bet13ISIT}, contents are stored in the users' local caches and in dedicated helper nodes distributed in the network. In contrast, our transmission scheme requires no helpers, but relies on multicast transmission to exploit content reuse.}. If D3 appears in queue 1 at the AP later, it will be dropped as it is already cached in user 1. Similarly, user 1 will not request for D3.

The main contribution of this paper is to introduce the opportunistic multicast transmission as an alternative to accommodate for more users in the network. This transmission scheme can also be used to replace the conventional multicast transmission if retransmission is essential: Instead of a single queue for the multicast group, multiple (virtual) queues can be set up to serve the users in group. Contrast to the study in \cite{Won09WC,Low10WC,Tsa11WC}, we have considered {\it multiple} multicast groups requesting for different contents. We also propose a multicast scheduling scheme for multiple-input multiple-output (MIMO) orthogonal frequency-division multiplexing (OFDM) systems with deadline constraints for real-time traffic. The proposed scheme also involves multicast precoding and user selection. User selection forms a multicast user group that consists of one intended user and multiple unintended users. Multicast regrouping, which is not considered in \cite{Won09WC}, is necessary due to the limitation of group size in practical network. We further investigate how to set protocol parameters to maximize the number of users accommodated in the network.

The rest of the paper is organized as follows. We describe system model and formulate a scheduling problem in Sections \ref{sec:sys} and \ref{sec:opt}, respectively. In Section \ref{sec:lyp}, we develop an algorithm based on the Lyapunov optimization. We provide simulation scenario and results in Section \ref{sec:sim} and conclude the paper in Section \ref{sec:con}.

\section{System Model} \label{sec:sys}

We first introduce a transmission procedure in a medium access control (MAC) layer, and then elaborate multicast precoding in a physical (PHY) layer.

\subsection{Transmission Procedure in MAC Layer}
The system operates in time slot with duration whose length is a bounded variable. The AP maintains $K$ queues, each of which supports a dedicated user, which implies there are $K$ users in the networks. Multiple data frames are allowed to be transmitted in each time slot as long as the transmission time does not exceed a given bound. Denoting a supported user set by $\mathcal{S}$, we introduce the detailed procedure in MAC layer.

{\bf Step 1}: At the beginning of each time slot, a user is selected by the scheduler to be served by AP. AP can operate in either unicast or multicast transmission. If the data frames to be transmitted are not required by other users (currently or in the future), unicast transmission (the cardinality of user set $|\mathcal{S}|=1$) is scheduled and accordingly single user MIMO precoding (beamforming in IEEE 802.11ac \cite{IEEE13AC}) is used. Otherwise, multicast transmission for multiple users ($|\mathcal{S}|>1$) is selected. The multicast precoding used will be designed in next subsection. For practical reasons, the maximum number of multicast users is limited to four, i.e., $|\mathcal{S}|\leq 4$. To balance the tradeoff between multiuser diversity and multicast gain, multicast regrouping is necessary \cite{Low10WC,Tsa11WC}. To select the multicast users, we use the norm criterion $\mathsf{E}_{n\in \mathcal{N}}\|\H_{n,k} \H_{n,s}^H\|_F^2$, where $n$ is subcarrier index and $\mathcal{N}$ is a subcarrier set; $\H_{n,k}$ and $\H_{n,s}$ is the channel gain matrix of subcarrier $n$ from AP to an intended user $k$ and an unintended user $s$. These values are sorted in descending order and the first three unintended users are selected to form the multicast group with the intended user $k$. Note that this multicast user grouping  is completely opposite to a multiuser (MU)-MIMO precoding. The link abstraction model for mapping the transmission mode to modulation and coding scheme (MCS) is based on mutual information approach given in \cite{Tan08JSAC}.
The scheduling priority is based on the head-of-line (HOL) delay, outdated transmission rate (due to outdated channel information), transmission  time required, frame length and the number of multicast users. The details on the design of this priority are deferred until the later sections.

{\bf Step 2}: Next, AP requests for channel state information (CSI) feedback from all users in $\mathcal{S}$. The channel sounding procedure in IEEE 802.11ac \cite{IEEE13AC} is applied. The AP sends a null data packet  announcement (NDPA) frame to notify the users to prepare for the channel measurement. The users measure the channel based on the null data packet (NDP) transmitted after the NDPA frame, followed by CSI feedback from one user to the AP. The AP then polls the remaining users for their respective CSI  if necessary.

{\bf Step 3}: Upon receiving the CSI, an MCS is selected for the transmission. The selection approach is the same as in Step 1. For the multicast transmission, the smallest MCS over all users is selected so as to ensure that all users can receive the data frames correctly, which will be used to design multicast precoding later. The number of data frames to be sent is determined by the MCS and the maximum allowable transmission time, which is also known as transmit opportunity (TXOP) in IEEE 802.11ac \cite{IEEE13AC}.

{\bf Step 4}:
After transmitting the frames, the AP also expects acknowledgement (ACK) frames from the users. After the first user has sent back the ACK frame, the AP sends ACK requests to the remaining users for their respective ACK if necessary. The ACK procedure is similar to the groupcast with retries (GCR) service in IEEE 802.11aa \cite{IEEE12AA}. After ACKs have arrived, the channel is released for contention.

{\bf Step 5}: Retransmission of erroneous frames is allowed. Only the erroneous packets for the intended user is retransmitted. In addition, retransmission has a higher priority than scheduling user for new transmission.

Summary of the transmission procedure at MAC layer:


\begin{framed}
\begin{enumerate}[Step~1:]
\item Select an intended user for transmission; Select (unintended) user(s) and form unicast or multicast user set; Estimate the MCS for transmission; Scheduling priority among users based on HOL delay, transmission rate and time, and packet length.
\item Request CSI feedback.
\item Determined MCS and number of frames to transmit based on current CSI feedback.
\item Transmit the packets and receive ACK/NACK from users.
\item Retransmission if it is necessary.
\end{enumerate}
\end{framed}

\subsection{Multicast Precoding in PHY Layer}

We consider a downlink {\it multicast} MIMO-OFDM system with one AP ($N_t$ transmit antennas) and $K$ users ($N_r$ receive antennas each), in which a common message is sent to all the users in $\mathcal{S}$ through $N$ subcarriers. Note that MU-MIMO transmitter sends individual message to each user. For a given time slot $t$, the scheduler at the AP selects a subset of users $\mathcal{S}$ to serve simultaneously. We assume that channel matrix $\H_{n,k}$ includes large- and small-scale fadings and is static during each transmission. For subcarrier $n\in\mathcal{N}=\{1,\ldots,N\}$, the $N_r \times 1$ received signal of user $k$ is given by
\beqna\nonumber
\r_{n,k} = \H_{n,k} \W_n \x_n + \z_{n,k},
\eeqna
where $\W_n\in\mathbb{C}^{N_t \times N_s}$ denotes the precoding matrix with the Frobenius norm equal to one, i.e., $\|\W_n\|_F^2=1$; $\x_n$ denotes the $N_{s} \times 1$ transmitted symbols with $N_{s}\leq N_r$ and $\mathsf{E}[\x_n\x_n^H] = \I_{N_{s}}/N_{s}$; and $\z_{n,k}$ denotes the additive Gaussian noise (AWGN) with zero mean and $\mathsf{E}[\z_{n,k}\z_{n,k}^H] = N_0 \I_{N_r}$, and the superscript $H$ represents the Hermitian transpose.


MU-MIMO precoding is typically designed to mitigate multiuser interferences so maximizing sum rate across all users is justifiable. In contrast, multicast MIMO precoding is designed to maximize the minimum sum rate \cite{JiLu06}, so that all users can receive the common message $\x$ as pointed out in Step 3 in previous subsection. Remind that the MCS is determined based on the minimum rate user. Thus, the multicast MIMO-OFDM precoding design problem is formulated as follows:
\beqi\label{optP}
\underset{\{\W_n\}}{\rm max.}\; \underset{k\in\mathcal{S}}{\rm min} &=&\!\left \{
{\sum_{n\in\mathcal{N}}} \!\log_2 {\rm det}\!\left(\! 1\!+\!  \frac{ {\bf H }_{n,k} {\bf W}_n{\bf W}_n^H {\bf H}_{n,k}^H}{N_0} \right)\!\!\right\}~~~~\inum\label{objo}\\
&{\rm s.t.}~&\|{\bf W}_n \|_F^2\leq1,~\forall n\in\mathcal{N},\inum\label{conso1}
\eeqi
where (\ref{conso1}) is for the transmit power constraint. Note that there is no multiuser interference and $\W_n$ is common for all users. This is critical difference between MU-MIMO and multicast. The upper bound of problem (\ref{optP}) can be efficiently obtained by using standard semi-definite progamming techniques.

However, for real-time scheduling of OFDM system, the optimization is computationally complexity-intensive. Instead of the optimal multicast precoding, we consider a near-optimal precoding based on a linear precoding principal, i.e., a precoding matrix $\W_n$ lies in the space spanned by $\{\H_{n,k}^H\}$, as follows \cite{JoNgTaSu04}:
\beqna
\W_n^* &=& \alpha\sum_{k\in \mathcal{S}} \Big\{ {\H_{n,k}^H}\U_{n,k}^H\Big/{\big\|\H_{n,k} \U_{n,k}^H\big\|_F^2} \Big\},~\forall n\in\mathcal{N}, \label{eqn:W}
\eeqna
where $\alpha$ is the normalization constant to fulfil $\|\W_n\|_F^2\leq1$; $\U_{n,k}=[{\bf u}_{n,k,1}\cdots {\bf u}_{n,k,N_s}]^H\in\mathbb{C}^{N_s\times N_r}$; and ${\bf u}_{n,k,i}$ is a left singular vector corresponding to the $i$th largest singular value of $\H_{n,k}$. When $N_s=N_r$, without loss of generality, we set $\U_{n,k}={\bf I}_{N_s}$ in (\ref{eqn:W}). The channel gain matrices are replaced by feedback channel estimates in this work.

\section{Scheduling Problem Formulation}\label{sec:opt}

Consider a finite number of time slot $T$, whose duration consists of the request for CSI feedback, the transmission of CSI, the transmission of data frames, the transmission of ACK frames, and the backoff period. Note that there is no contention because we do not consider any uplink traffic.

Our goal is to find a scheduling scheme that makes binary transmission decisions $\mu_k[t]$ and frame dropping decisions $\omega_k[t]$ for each time slot of duration $T[t]$. A decision of one implies that positive action is taken. We denote the amount of bit to be transmitted as $b_k[t]$ and the amount of bit to be dropped as $d_k[t]$. The $b_k[t]$ values are determined by $\mu_k[t]$ and current CSI $\eta_k[t]$, while the $d_k[t]$ values are determined by $\omega_k[t]$. Since we do not consider MU-MIMO in this work, we have orthogonal channel transmission where $\sum_k \mu_k[t] \le 1$. In addition, we also constrain $b_k[t]$ and $d_k[t]$ such that no frame fragmentation is required. The dynamics of the queue is modeled as follows:
\beqna
Q_k[t+1] &=& \max\{0,Q_k[t] - b_k[t] - d_k[t]\} + A_k[t], \label{eqn:Q_kt}
\eeqna
where $A_k[t]$ is the amount of bits arrived at time slot $t$. Note that data frames arriving at the current time slot will only be served at the next time slot.

We design our scheduling scheme to maximize the transmission rate and minimize the dropping rate. Hence,the optimization problem is formulated as follows:
\beqna
\max_{\mu_k[t],\omega_k[t]}\,\, \frac{\sum_k \overline{b_k} - v_k \overline{d_k} }{\epsilon \overline{T}}, ~\forall k,\label{eqn:opt1}
\eeqna
where $v_k$ and $\epsilon$ are the parameters for a maximum deadline constraint and a measuring unit for HOL delay, respectively.Here, we define $\overline{b_k}$ as the  time average of transmitted bit $b_k[t]$ as
\beqna
\overline{b_k} = \frac{1}{T} \sum_{t=0}^{T-1} b_k[t],
\eeqna
and $\overline{d_k}$ and $\overline{T}$ represent the time average of $d_k[t]$ and $T[t]$, respectively.

We use Lyapunov optimization theory \cite{Nee10BK,Nee10Arx} to design scheduling scheme for arbitrary $\eta_k[t]$ and $A_k[t]$. The decision vectors $\bmu [t] = [\mu_1[t],\dots,\mu_K[t]]$  and $\bomega[t] = [\omega_1[t],\dots,\omega_K[t]]$ are chosen by minimizing an upper bound on a drift-plus-penalty ratio \cite{Nee10Arx}, which will be defined later. At each time slot, we need to solve a quasiconvex problem. To reduce the complexity, we reformulate the optimization problem as
\beqna
\min_{\mu_k[t],\omega_k[t]}\,\, \epsilon \overline{T} - \beta  \sum_k \overline{b_k}
+ \beta  \sum_k v_k\overline{d_k},~\forall k,  \label{eqn:opt2}
\eeqna
where a given parameter $\beta$ has been added. Note that if $1/\beta$ is the maximum of \eref{eqn:opt1}, the problem \eref{eqn:opt2} is equivalent to \eref{eqn:opt1}.
\begin{rem}
In the above formulation, we assume that the scheduler has the current CSI, which is not true for the transmission procedure described previously. Though outdated CSI is available at the point of making the scheduling decision, the assumption makes the formulation more concise. It is also assumed that the transmission is error-free; therefore, retransmission is unnecessary. If channel error is incurred, we can consider the expectation of $b_k[t]$, $d_k[t]$ and $T[t]$ over the error events. For multiple data frames  and retransmission attempts, there are no close-form solutions for these expectation terms. Hence, we devise a heuristic scheduling scheme that behaves properly if the MCS is selected with a low probability of a channel error event.
\end{rem}

\section{Proposed Scheduling Scheme} \label{sec:lyp}
We propose a heuristic scheduling scheme having sequential structure with transmission decision and frame  dropping decision. Since we do not consider the overflow of queues, it is a better strategy to serve and then drop the remaining frames.
Let $Z_k[t]$ represent the HOL delay at time slot $t$ and $\widetilde{Z}_k[t+1]$ is an intermediate update on HOL delay after $\mu_k[t]$ has been made. We define 
\beq
\phi_k[t] = \phi_k(\omega_k[t])\}\triangleq
\begin{cases}
\min(M_k[t],\widetilde{Z}_k[t+1]), & \textrm{if $\omega_k[t] = 1$},\\
0, & \textrm{otherwise}.
\end{cases}
\eeq
To obtain $Z_k[t+1]$, $\widetilde{Z}_k[t+1]$  is reduced by the inter-arrival time between the HOL frame and the subsequent frame $M_k[t]$ if the queue is not empty after frames are dropped. If the queue becomes empty, it is reduced by $\widetilde{Z}_k[t+1]$. However, if no frame is dropped, $Z_k[t+1] = \widetilde{Z}_k[t+1]$. Hence, $Z_k[t+1]$  is updated as 
\beqna
Z_k[t+1] &= &\max \{0,\widetilde{Z}_k[t+1] - \phi_k(\omega_k[t])\}. \label{eqn:H1}
\eeqna
Similarly, $\widetilde{Z}_k[t+1]$ is given by
\beqna
\widetilde{Z}_k[t+1] &= & \max \{0,Z_k[t] - \psi_k(\mu_k[t])\} \label{eqn:H2},
\eeqna
where  $\psi_k(b_k[t])$ is given by
\beq \psi_k[t] = \psi_k(\mu_k[t]) \triangleq
\begin{cases}
\min(M_k[t],Z_k[t]), & \textrm{if $\mu_k[t] = 1$},\\
-\epsilon T[t], & \textrm{otherwise}.
\end{cases}
\eeq
In this work, we set $\epsilon = 1,000$, and hence, the HOL delay in \eref{eqn:H1} and \eref{eqn:H2}  are measured in milliseconds.

Defining the quadratic Lyapunov function
\beqna
L[t]  &\triangleq&   \frac{1}{2}\sum_k Z_k[t]^2 \nn
\eeqna
and the Lyapunov drift on slot $t$ as $\Delta[t] \triangleq L[t+1] - L[t]$, the algorithm is designed to minimize a bound on the following drift-plus-penalty ratio expression \cite{Nee10Arx}:
\beqna
\Delta[t] + V\left\{\epsilon T[t] - \beta  \sum_k b_k[t]
+ \beta  \sum_k v_kd_k[t]  \right\}, \label{eqn:dpp}
\eeqna
where $V\ge 0$  is a control parameter chosen for performance tradeoff. The Lyapunov drift $\Delta[t]$ is upper bounded as shown in Lemma 1.
\begin{lem} $\Delta[t]$   satisfies
\beqna
\Delta[t]
&\le& B - \sum_k Z_k[t] \psi_k[t] - \sum_k \widetilde{Z}_k[t+1]) \phi_k[t], \label{eqn:dbound}
\eeqna
where $B$ is a finite constant.
\end{lem}

Bounding \eref{eqn:dpp} with \eref{eqn:dbound} requires $M_k[t]$, which is a random variable whose value is known only after the scheduling decisions are made. Hence, we approximate $\phi_k[t]$ and $\psi_k[t]$ by $\widetilde{\phi}_k[t]$ and $\widetilde{\psi}_k[t]$, respectively. We define  $\widetilde{\phi}_k[t] = \widetilde{Z}_k[t+1]$ if $\omega_k[t] = 1$ and $\widetilde{\psi}_k[t] = Z_k[t]$ if $\mu_k[t] = 1$.
To minimize this upper bound, the drift-plus-penalty scheme determines the values of $\mu_k[t]$ and $\omega_k[t]$ decisions every time slot. We label this scheme as a Lyapunov optimization (LO) scheduler and summarize it as follows:

\begin{framed}
\begin{enumerate}[Step~1:]
  \item \underline{Scheduling}: For each time slot $t$, choose $\mu_k[t]$ to
\beqna
\hspace{-8mm}
&&\max _{\mu_k[t]}\,\sum_k Z_k[t] \widetilde{\phi}_k[t] - V\epsilon T[t]
 + V \beta \sum_k b_k[t] \label{eqn:op_sch}
\eeqna
\item \underline{Frame Dropping}: For each time slot $t$, choose $d_k[t]$ to
\beqna
\hspace{-6mm} &&\max \,\, \widetilde{Z}_k[t+1] \widetilde{\psi}_k[t]
- V v_k\beta d_k[t] \label{eqn:op_drop}
\eeqna
\item \underline{Queues Updates}: Update the queues $Q_k(t)$, $Z_k(t)$ and $\widetilde{Z}_k[t+1]$ according to \eref{eqn:Q_kt}, \eref{eqn:H1} and \eref{eqn:H2}, respectively.
\end{enumerate}
\end{framed}

If user $k'$ is selected to be served, the objective function in \eref{eqn:op_sch} is given by
\beqna
Z_{k'}[t]^2 - \left(\sum_{k\neq k'} Z_k[t]  +V \right) \epsilon T_{k'}[t] +
 V \beta b_{k'}[t],
\eeqna
where $b_{k'}[t]$ is the maximum number of bits which can be transmitted while its corresponding transmission time $T[t] = T_{k'}[t] $ is still less than the predetermined threshold $T^{\max}$. As $\sum_k \mu_{k}[t] \le 1$ for orthogonal channel transmission, the scheduling problem can be further decomposed into
\beqna
k' &=& \arg \max_k Z_{k}[t]^2 - \left(\sum_{j\neq k} Z_j[t]  +V \right) \epsilon T_{k}[t] +
 V \beta b_{k}[t]. \nn
\eeqna
If the AP is multicasting the data frames to the $|\mathcal{S}'|$ users, the value of $b_{k'}[t]$ is increased by $|\mathcal{S}'|$ fold. As usual, the scheduling criterion includes the HOL delay and the transmission rate of the users. In addition, it also includes the transmission time and the number of bits transmitted.

The constraint set for $d_k[t]$ is given by $\{0,L_k[t]\}$, where $L_k[t]$ is the amount of bits (restricted to integer number of frames) that can be dropped from the queue. Solving \eref{eqn:op_drop}, we have
\beq
d_k[t] =
\begin{cases}
L_k[t], & \textrm{if $\widetilde{Z}_k[t+1]^2 \ge V\beta v_kL_k[t]$},\\
0, & \textrm{otherwise}.
\end{cases}
\eeq

\subsection{Deterministic Performance Bound}

It can be shown that the drift-plus-penalty scheme described comes within $O(1/V )$ of the utility of a genie-aided $T'$-slot lookahead algorithm with an average delay constraint of $O(V )$ \cite{Nee10Arx}. Furthermore, we can ensure that the frames are dropped with a worst delay given in the following Lemma 2.
\begin{lem} \label{lem:delay}
Suppose that $L_k[t] \le L^{\max}$  is the minimum number of bits to be dropped (restricted to integer number of frames) such that the HOL delay of the subsequent frame has been decreased by more than $T^{\max}$. Then frames are dropped with a maximum value of $Z_k^{\max} = \sqrt{Vv_k\beta L^{\max}} \ge Z_k[t]$.
\end{lem}
\begin{IEEEproof}
The proof is shown via induction. Suppose $Z_k[t] \le Z^{\max} - \epsilon T^{\max}$, this implies $\widetilde{Z}_k[t+1] \le  Z^{\max}$ from \eref{eqn:H1} and subsequently $Z_k[t+1] \le  Z^{\max}$ from \eref{eqn:H2}. Now suppose $Z^{\max} - \epsilon T^{\max} < Z_k[t] \le Z^{\max}$, we have $ Z^{\max}  < \widetilde{Z}_k[t+1] \le  Z^{\max} + \epsilon T^{\max}$  from \eref{eqn:H1}. By design, frame dropping occurs and then $Z_k[t+1] \le  Z^{\max}$.
\end{IEEEproof}

\begin{rem}
The above scheduling scheme does not consider retransmission. Hence, Lemma \ref{lem:delay} no longer holds in this context. An additional mechanism is needed for dropping frames after the deadline. The behavior of the scheduling scheme is explored via simulation in the next section.
\end{rem}

\section{Simulation Results}\label{sec:sim}

\begin{table}[!t]
\centering

\resizebox{.48\textwidth }{!}{
\begin{threeparttable}
\caption{Simulation Parameters (from IEEE 802.11ac Network).}
\label{tab:s0}
\begin{tabular}{|c|c|c|}\hline
\bfseries Parameters & \bfseries Value \\
\hline\hline
Number of contents  &	10    \\\hline
Data frame size  &	1,000 Bytes  \\\hline
Traffic load    &	0.5, 1, 2 , 5 Mbps  \\\hline
Frame exchange sequence & CSI+Data+ACK+DIFS\tnote{$\diamond$}\; + Backoff \\\hline
Max transmission time  &  3 msec\\\hline
Max retransmission attempts &  3\\\hline
$N_t,~N_r,~N_{s}$, & 4,~1,~1 \\\hline
SNR  & 12--45 dB\\\hline
Bandwidth  & 20 MHz\\\hline
MCS & 0--7 \\\hline
Deadline $Z^{\max}$  & 200 msec\\\hline
Channel model & D \cite{IEEE0411N} \\\hline
Link abstraction model & Based on mutual information \cite{Tan08JSAC} \\\hline
\end{tabular}
{\footnotesize
\begin{tablenotes}
\item[$\diamond$] {DIFS: Distributed Coordination Function (DCF) Interframe Space.}
\end{tablenotes}}
\end{threeparttable}}
\end{table}

\subsection{Simulation Framework}

\begin{figure}[!t]
\begin{center}
\epsfxsize=0.45\textwidth \leavevmode
\epsffile{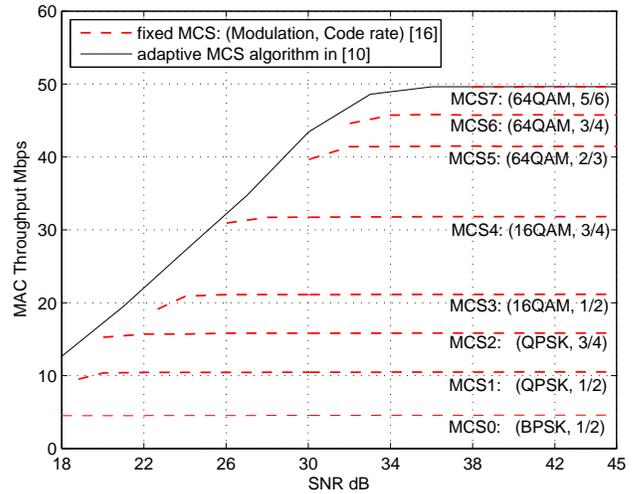}
\caption{Evaluation of MAC throughput performance over average SNR.}
\label{fig:p1}
\end{center}
\end{figure}

Table \ref{tab:s0} lists the simulation parameters used in the IEEE 802.11ac system simulator. The simulator is based on a single cell layout. The link abstraction model is based on the mutual information approach given in \cite{Tan08JSAC}. In Fig. \ref{fig:p1}, the average MAC throughput is plotted against average signal-to-noise ratio (SNR) for the $N_t = 4$, $N_r = 1$ and $N_s = 1$ system in channel D. The dashed lines correspond to the throughput of fixed MCSs. The solid line corresponds the throughput achieved by our link adaptation algorithm \cite{Tan08JSAC}. We observe the operating SNR range of this system varies from 18 to 45 dB. We therefore consider the following two deployment scenarios:
\begin{itemize}
\item Case 1: 18 dB $\le$ SNR $\le$ 45 dB
\item Case 2: 30 dB $\le$ SNR $\le$ 45 dB
\end{itemize}
Case 1 attempts to cover the whole operating SNR range, while Case 2 looks at the high SNR region. For both cases, the SNR of the users are uniformly selected from their respective ranges.

For traffic model, we consider each user requests for one of the 10 different contents and the selection is done randomly. The start of the frame arrival to the queue is also randomly with the interval of 500 ms. The frames arrive from constant bit rate flow for 2 sec and then pauses for 1 sec. The size of the frame is set to 1,000 Bytes. This cycle is repeated until the simulation is ended. The time duration of each simulation run lasts 30 sec and all simulation results are averaged over 100 sessions.

We now discuss the parameter selection for the LO scheduler. From Fig. \ref{fig:p1}, the maximum throughput is around 50 Mbps. We select the estimated  throughput for  LO scheduler to be 25 Mbps and hence $\beta = 4\times 10^{-5}$. Note that the performance of LO scheduler is not sensitive to the value of estimated  throughput as long as the estimated  throughput is of the same order as the simulated throughput. The maximum HOL delay of all users is set to $Z^{\max} = 200$ ms and $L^{\max} = 8,000$ bits. This implies that $Vv_k = 1.25\times 10^{5}$. We vary the value of $V$ from 1 to 10,000 and found that the $V = 1,000$ gives the best performance. Therefore, we set $v_k = 125$.

\subsection{Performance of the Schedulers}

Since the traffic load (and hence, the arrival rates) are fixed in the simulation, we consider the  number of users that can be supported by the system as our performance metric. This user capacity depends on the the traffic load and the choice of the outage criteria. The outage criteria in our simulation is similar to the evaluation methodology in \cite{3GPPTR25.814}. We consider a user to be in outage if more than 1\% of the frames are either lost or delivered with a delay exceeding the de-jitter buffer delay. The system is considered to be in outage if more than 1\% of the users are in outage. Table \ref{tab:s3} lists the user capacities for Case 1 and Case 2 for various schedulers and traffic load.

We first look at the user capacities for the schedulers without multicast transmission. As shown in Table \ref{tab:s3}, the user capacity for traffic load of 0.5 Mbps in Case 1 are 57, 39, and 32 for LO,  Maximum-Largest Weighted Delay First (MLWDF)  \cite{Sto01AP}, and Round Robin (RR) schedulers, respectively. For Case 1, the LO scheduler has the highest capacity, while the RR scheduler scheduler has the lowest capacity. The lack of transmission rate and HOL delay in the calculation for scheduling priority is the reason why RR has the worst capacity. The addition of transmission time in the calculation for scheduling priority allows the LO scheduler to achieve higher capacity than an MLWF scheduler. The gain is more significant for high density deployment with low data rate. For traffic load of 0.5 Mbps, LO scheduler can support up to 57 users, compared to 39 users for  MLWF scheduler. That implies a gain of 46$\%$. However, the gain vanishes if the SNR of the users are very high as shown in Case 2. From Table \ref{tab:s3}, we see that the LO scheduler has similar user capacity as the MLWF scheduler, yet it still outperforms the MLWF scheduler.

Next, let look at the user capacities for the schedulers with multicast transmission. As shown in Table \ref{tab:s3}, the user capacity for Case 1 and traffic load of 0.5 Mbps are 67, 53, and 41 for LO, MLWF, and RR schedulers, respectively. The gain from multicast  is more significant for high density deployment as the opportunity for multicast increases as the number of users increases. For data rate of 0.5 Mbps in Case 1, the LO scheduler can support up to 57 users and 67 users for unicast and  multicast mode, respectively. That translates to an increase of 17$\%$ in user capacity. Higher gain can be achieved by using multicast mode with MLWF and RR schedulers, although their user capacities are still lower than that of the proposed LO scheduler. In addition, the gain is more significant if the SNR of the users are very high as shown in Case 2. For data rate of 0.5 Mbps, the user capacity of  LO scheduler increases from 53 to 72, which is a gain of 35$\%$.

Finally, we look at the size of cache required at the user's device in Table \ref{tab:s5}. In general, the higher the data rate, the larger the size of cache required. The opportunity for multicasting also determines the size of cache. For a given traffic load, users in Case 2 require larger size of cache than users in Case 1.

\begin{table}[t]
\caption{Capacities for Case 1 (unit: number of users).}
\label{tab:s3}
\centering
\setlength{\tabcolsep}{.4em}
\resizebox{.48\textwidth }{!}
{
\begin{tabular}{|c|c c|c c|c c||c c|c c|c c|}
\hline
\multirow{2}{*}{\bfseries Scheduler} & \multicolumn{6}{c||}{\bfseries Data Rate: Case 1}&\multicolumn{6}{c|}{\bfseries Data Rate: Case 2} \\ \cline{2-13}
                           & \multicolumn{2}{c|}{0.5 Mbps}    & \multicolumn{2}{c|}{1 Mbps}   & \multicolumn{2}{c||}{2 Mbps}           & \multicolumn{2}{c|}{1 Mbps}    & \multicolumn{2}{c|}{2 Mbps}   & \multicolumn{2}{c|}{5 Mbps}          \\ \hline
multicast                  & no & yes      & no & yes  & no & yes        & no & yes      & no & yes  & no & yes        \\\hline\hline
LO                         & 57&67         & 29&31         & 12&12      & 53&72        & 27&36        & 10&13       \\ \hline
MLWF                       & 39&53         & 21&26         & 10&11     & 49&69        & 26&33        & 10&12       \\ \hline
RR                         & 32&41         & 17&20         &  8&9      & 27&51        & 19&29        & 10&11       \\ \hline
\end{tabular}
}
\end{table}


\section{Conclusion}\label{sec:con}

We have proposed a transmission scheme for exploiting content reuse with opportunistic user requests. The proposed opportunistic multicast transmission is considered in a unicast environment to reduce the wireless resource usage. The Lyapunov optimization approach for the multicast scheduling scheme is designed for real-time traffic. Numerical simulations over WLAN networks have been presented to show the effectiveness of the proposed scheme. It is observed that significant multicast gain (35$\%$) is achievable at higher operating SNR environment with the expense of larger cache memory at user's device.

\begin{table}[t]
\caption{99 percentile cache size for multicast (unit: frames).}
\label{tab:s5}
\centering
\setlength{\tabcolsep}{.4em}
\resizebox{.28\textwidth }{!}
{
\begin{tabular}{|c|c|c|c|c|c|c|}
\hline
\bfseries Scenario                   & \multicolumn{3}{c|}{\bf Case 1} & \multicolumn{3}{c|}{\bf Case 2} \\ \hline
\multirow{2}{*}{\bfseries Scheduler} & \multicolumn{6}{c|}{{\bf{Data Rate}} Mbps}                 \\ \cline{2-7}
                           & 0.5     & 1       & 2       & 1       & 2       & 5       \\ \hline \hline
LO                         & 27   & 55   & 80  & 68   & 126  & 292  \\ \hline
MLWF                       & 26   & 53   & 77   & 60   & 114  & 253  \\ \hline
RR                         & 29   & 53   & 80  & 65   & 127  & 267  \\ \hline
\end{tabular}
}
\end{table}

\vspace{.7cm}

%

\end{document}